# Administration Costs in the Management of Research Funds; A Case Study of a Public Fund for the Promotion of Industrial Innovation

David R Walwyn
Correspondence to david.walwyn@up.ac.za

*Department of Engineering and Technology Management, University of Pretoria, South Africa*
(Dated: September 19, 2016)

**Abstract**

Research funding agencies routinely use a proportion of their total revenues to support internal administration and marketing costs. The ratio of administration to total costs, referred to as the administration ratio, is highly variable and within any single fund depends on many factors including the number and average size of projects and the overall efficiency of the funding agency. Surprisingly there is little published information on the administration ratio and as a result little guidance for fund managers concerned about the optimisation of their research portfolios. In this study, the standard agency activities have been identified and used to develop a model of administration costs vs. expected outcomes. In particular, the model has been designed to estimate the optimum portfolio success rate and administration ratio as a function of a range of key input variables including the project size, the complexity of proposal evaluation and project management, the risk tolerance of the funder and the targeted research domain. The predicted administration ratio is then compared to actual data as obtained from a detailed analysis of the Technology for Human Resources in Industry Programme. The model correctly identifies the reason for the programme's 50% increase in output per unit of funding over the period 2009 to 2014, despite a 20% drop in funding (in real terms) and an increase in the administration ratio. This result suggests that the model could be used more widely to both benchmark organisational efficiency and improve the overall administration of research funds.

## 1. INTRODUCTION

Funding agencies, which manage public research and development (R&D) funds on behalf of wholesale funders, are a pervasive feature of all national systems of innovation (1). Such agencies routinely use a proportion of their total fund revenues to support internal administration costs. The ratio of fund administration costs to total revenues, referred to in this article as the administration ratio (AR), is highly variable and depends on many factors other than the overall efficiency of the responsible funding agency. Illustrative values for the ratio are given in Table 1; it is apparent that AR varies considerably with no clear pattern between the various organisations.

**Table 1. Illustrative values for the administration ratio of funding agencies**

| Organisation | Country | Year | AR | Source |
|---|---|---|---|---|
| Technology Innovation Agency | South Africa | 2012 | 39% | Andreas Bertoldi, David Gardner (2) |
| Cancer Research UK | United Kingdom | 2013/14 | 31% | Cancer Research UK (3) |
| Water Research Commission | South Africa | 2009 | 22% | Ronald David Grace (4) |
| British Heart Foundation | United Kingdom | 2013 | 19% | British Heart Foundation (5) |
| National Endowment for Democracy | United States | 2013 | 16% | National Endowment for Democracy (6) |
| Support Programme for Industrial Innovation | South Africa | 2010/11 | 17% | Support Programme for Industrial Innovation (7) |
| Wellcome Trust | United Kingdom | 2011/12 | 11% | Annual Report 2013 |
| National Research Foundation (NRF) | South Africa | 2013/14 | 9.4% | National Research Foundation (8) |
| Technology and Human Resources for Industry Programme (THRIP) | South Africa | 2013/14 | 5.8% | THRIP (9) |
| Research and Innovation Support and Advancement | South Africa | 2013/14 | 5.3% | National Research Foundation (8) |

Funders and researchers generally respond energetically to the debate about the funding of administration and overhead costs in a funding agency. Funders, not without reason, see the diversion of funds from their intended target (the researchers) as a tax on their budgets, and may not appreciate the link between effective, although often costly, administration and portfolio outcomes. Researchers on the other hand resent the reduction in their project budgets as a consequence of the agency's expenses, the latter seemingly funding a whole slew of unnecessary activities.

These sentiments are also echoed in the literature on charities, non-governmental organisations and philanthropic entities (collectively referred to in this analysis as NGOs), where donors are generally loath to fund overhead or fundraising costs. In a survey undertaken by Grey Matter Research & Consulting, it was revealed that "the average American believes a charity should spend no more than 23% on overhead but charities actually spend 36%" and that "62% of Americans believe the typical charity spends more than it should on overhead" (10).

On the other hand, many NGOs feel that the issue is poorly understood and that the percentage of NGO expenses that are used for administrative/fundraising costs, collectively referred to as overhead, is a poor measure of a NGO's performance, although commonly used (11). Donors are advised to subscribe to an NGO's results, rather than its expenses, and trust them to make the right decisions on how to spend your money (12). Despite this advice, it appears that NGOs generally respond to the unrealistic expectations of their funders by under-financing and under-reporting overheads, leading eventually to the Nonprofit Starvation Cycle (13). The way forward for such organisations, according to Megan Roesner (14), is to source funding only from donors who understand the importance of investing in infrastructure and people.

Surprisingly there is little published information on AR and hence little guidance for fund managers in the optimisation of their research portfolios. The objective of this analysis has been to understand and explain these differences, and to separate legitimate expenditure from organisation inefficiency, where the latter is generally not justifiable and should be avoided. A model for the determination of an optimum AR based on the fund's intrinsic characteristics and the funder's risk tolerance has been developed. The inputs for the model have been calibrated using published information on project success rates and data from a range of local funding agencies. In the final section, the historical performances of a South African funding instrument, known as the Technology and Human Resources for Industry Programme (THRIP), have been compared against the model's prediction. The potential insight and the limitations of the model, including relevant source or calibration data, are then discussed. In the final section, recommendations to improve the achievement of the optimum administration ratio and to establish a common perspective on the necessary, but often resented, cost of administration are made.

## 2. METHODOLOGY

The research has followed a quantitative methodology with the initial development of a predictive model for administration costs as a function of the designated input variables, followed by validation of the model in the case study of THRIP. The first step in the definition of the model was to itemise all the components of the administration costs and allocate approximate costs to each activity, supported by primary data obtained from a funding agency. The components of the total administration cost were specified as follows:

- all direct costs including the salary costs for personnel who work directly on the project, office expenses (telephone, stationary, consumables), funding of reviewers and external consultants, direct depreciation, marketing and training
- all indirect costs such as computer services, security and auditing, proportioned according to the total revenue for the funding unit relative to the overall organisational revenue.

The separate tasks together with their estimated costs are listed in Table 2. It is noted that the latter values are indicative only and will vary considerably between agencies and instruments.

The principle of discretionary tasks forms the core of the analytical approach in this study. Clearly there are a number of tasks which are routine and non-discretionary such as calls for proposals, reports to funders and financial reports. However there are also a number of non-routine activities (such as external and detailed ex-ante or ex-post evaluation of proposals, regular project monitoring and detailed site visits) which add a level of confidence, but also cost, to the management of the portfolio. These activities increase the administration cost but simultaneously improve the project success rate. The trade-off between these two important consequences is discussed in more detail in Section 5.

In a comparative study of this nature, there are two important limitations which need to be considered, namely the lack of accounting agreement on the allocation of expenses and secondly the large variation in sources of revenue between different agencies. As a result, a direct comparison based on published financial information is difficult, since different accounting policies may have been followed in the compilation of the figures. In this study, a clear distinction is made between the funds that are transferred to the research projects and the expenses of the agency in the disbursement of such funds; the ratio of these two quantities forms the essence of the analysis.

**2.1 Definition of the Independent Variables**

The following independent variables are normally defined by the funder when the funding instrument is initially established:

- total value of the fund, referred to as $V_P$, usually defined by the funders on a 3 to 5 year basis e.g. the amount to be transferred to the funding agency will be US $10 million per year for 5 years, after which point the impact of the fund will be evaluated and its future determined
- approximate value ($V_i$) and duration of the individual projects e.g. each project will be funded for a three year period at about $0.5 million per year, giving a project budget of $1.5 million
- the funder's preference in respect of portfolio optimisation, being either to achieve the optimum portfolio success rate (PortSR) irrespective of the associated administration cost;
- or to minimise the portfolio risk, where this is expressed as a minimum expected project success rate ($PSR_{min}$). The latter is an important but also the most difficult to define of the various input variables and a separate section is devoted to a discussion of this parameter (see Section 2.2).

**Table 2. Typical administration tasks undertaken by funding agencies**

| Component | Units | Indicative Value (ZAR1000's)[1] | Description and Comments |
|---|---|---|---|
| Tasks common to all funds (non-discretionary) | % of portfolio value | 5% | Prepare budgets and motivation for funding, report to funder, terms of reference, call for proposal, collation, preparation of evaluation matrix, internal screening, notice of award and project contracting |
| Detailed internal ex-ante evaluation | ZAR1000s/ project | 50 | More detailed ex-ante evaluation by agency |
| External ex-ante evaluation | ZAR1000s/ project | 400 | Development of expertise database, circulation of proposals, adjudication and financial support for adjudicators |
| Life cycle project monitoring | ZAR1000s/ project | 300 | Life cycle project monitoring |
| Internal ex-post evaluation | ZAR1000s/ project | 150 | Internal ex-post evaluation |
| External ex-post evaluation | ZAR1000s/ project | 600 | External ex-post evaluation and impact assessment |
| Other (Awardee Training) | ZAR1000s/ project | 200 | Develop community of practice, training of awardees, etc. |
| **Total (Maximum)** | | **1,700** | Value excludes the tasks common to all funds |

---

[1] One Rand (ZAR) = USD 0.07

In addition to the funder-defined input variables, there are two other important variables which are defined as follows:

- the level of technical risk for the investment area, which determines the intrinsic project success rate ($PSR_i$). This variable is covered in more detail in Section 2.3. $PSR_i$ is itself a function of both the technical risk of the knowledge domain, such as biotechnology, manufacturing, advanced materials or information and communication technologies; and the fund's focus in terms of the research, development and innovation value chain (RDI) such as basic research, applied research or human capital development.
- the level of organisation efficiency (ET), which covers the general performance of a funding agency organisation in the management of R&D funding. In the contest about who funds organisational overheads, this variable is at the core of the debate; further details are given in Section 2.4.

The relationships between the key variables as listed in Section 2.1 and used throughout this analysis are shown in Figure 1.

**Figure 1. Input variables and algorithm for the funding model**

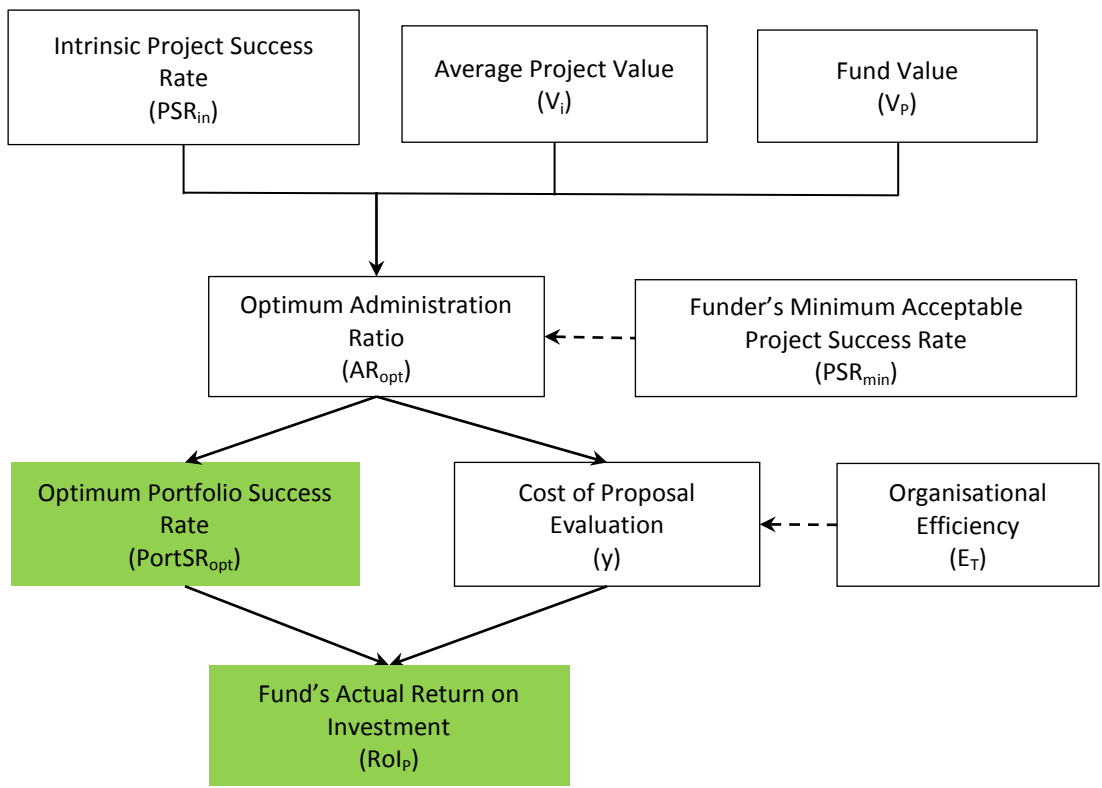

## 2.2 Risk Tolerance of the Funder

Funder's tend to be risk averse and expect high rates of project success. On the other hand, they also often fail to appreciate that risk mitigation implies higher levels of portfolio administration and hence higher cost relative to $V_P$. The trade-off between portfolio success and project success is core to this analysis; higher levels of project success can be achieved by increased investment in project review and oversight, but this is obtained at a cost to the overall portfolio metrics.

It is hypothesised that some funders are able to express their level of risk tolerance in the form of a minimum acceptable PSR ($PSR_{min}$). The difference between $PSR_i$ and $PSR_{min}$, referred to as $\Delta PSR$, becomes a key input to the model and directly determines the complexity and hence cost of the evaluation and monitoring processes.

Alternatively, it is assumed that funders will generally be unaware of their risk preferences but will instead prefer to select an option of the optimum administration ratio ($AR_{opt}$) which delivers the optimum portfolio success rate ($PortSR_{opt}$). An algorithm for the determination of $AR_{opt}$ is explained in Section 3.3; at this point it is noted that this calculation is mathematically more complicated than the statement of the minimum preferred risk and depends on defining the quantitative relationship between $\Delta PSR$ and AR. Two options for the latter are presented, namely a linear correlation and a logistics function (see Section 3.3).

### 2.3 Technical Risk for the Investment Area

The project success rate (PSR), defined as the number of projects within the total portfolio which achieve their stated objectives, is determined by the combined influence of many variables including the competence of the researchers, good fortune, enduring hard work, the technical difficulty of the technology domain within which the project is located (e.g. software, biotechnology and optics), the level of maturity of the underlying technology and the availability of key research facilities. In this analysis, all of the exogenous variables (i.e. those variables over which the funding agency has no control) are ignored, and only the funding instrument's target with respect to the components of the research and innovation (R&I) value chain (e.g. human capital development, basic research, applied research, experimental development, commercialisation and operations) and the level of technical risk in the technology domain are considered as valid inputs to the model.

Funders generally express a desire to fund projects at specific points in the R&I value chain (such as human capital development or applied research) and in certain technology domains (such as pharmaceuticals), sometimes unaware of the greater level of risk that may be attached to such choices. Typical or average risk factors associated with either the technology domain or the value chain, or the variance of these average values, are not published although they may be collected by funding agencies. In this analysis, literature values published by David R Walwyn, David Taylor (15) for the average intrinsic project success rates ($PSR_{in}$) by technology domain and value chain focus have been used (see Figure 2). Examples of high and low risk areas are pharmaceuticals/biotechnology are software development respectively.

### 2.4 Organisational Efficiency

Finally in this section the issue of organisational efficiency is discussed. Thus far it has been assumed that the funding agency is 100% efficient, and that the administration cost is determined by the intrinsic risk profile of the portfolio and the funder's risk tolerance. Although this may define the administration frontier, it ignores levels of inefficiency which may be present within the agency and include expenditure on non-core activities.

It is these inefficiencies which funders wish to avoid, as noted in the introduction. Whilst key functions such as adjudication and financial management are important, unnecessary expenditure on marketing or information technology warrants censure from funders and researchers alike. Unfortunately it has not previously been possible to determine levels of organisational efficiency ($E_T$) since information on overhead costs as a function of instrument type and funder expectations has not been collected. The intention of this work has been to propose a universal methodology which could be used of determining the efficient PortSR, and hence $E_T$.

**Figure 2. $PSR_{in}$ as a function of technology domain and RDI focus**

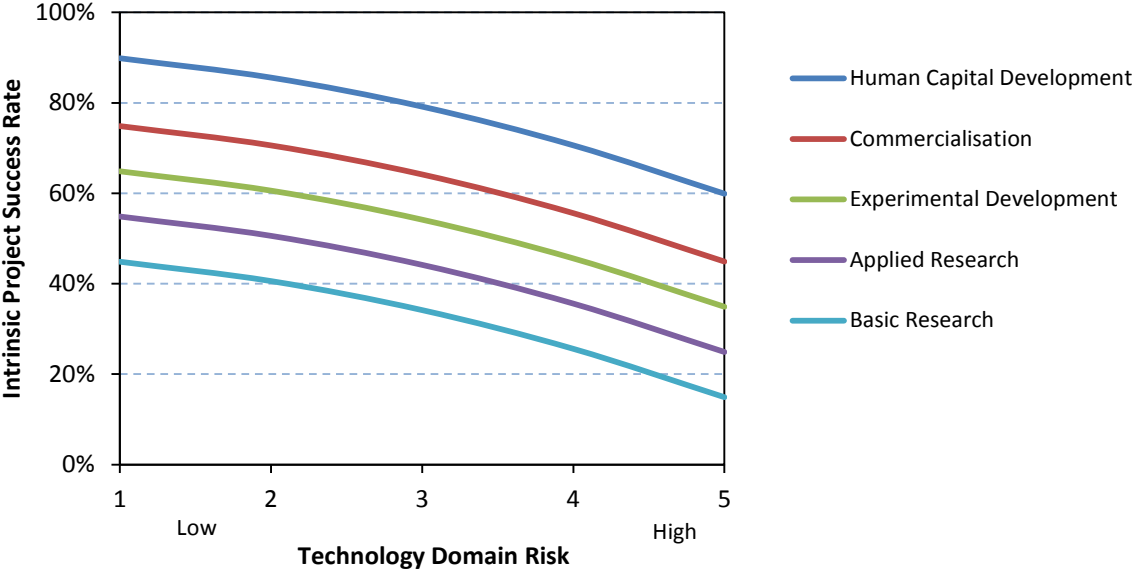

## 3. MATHEMATICAL EXPRESSION OF THE MODEL

### 3.1 Step One – Funder Input and Evaluation Complexity

The following information is required from the funder (example values are included in brackets):

- technology domain (e.g. biotechnology)
- RDI focus (e.g. experimental development)
- average project size, $V_i$ (e.g. $10 million)
- total fund value, $V_p$ (e.g. $0.5 million).

Based on the example values for RDI and technology domain, $PSR_{in}$ is obtained from Figure 2 as 54%.

### 3.2 Step Two – Cost of Administration and Successful Projects

The next step is to calculate the cost of administration. The relevant equations are:

$$Admin\ Cost = B * V_P + NP * y \qquad \ldots\ldots\ldots.. 1$$

$$NP = \frac{V_P - Admin\ Cost}{V_i} \qquad \ldots\ldots\ldots\ldots 2$$

$$Admin\ Ratio\ (AR) = \frac{Admin\ Cost}{V_P} \qquad \ldots\ldots\ldots\ldots 3$$

where:

    $y$          = average evaluation cost per project (R million/project)
    $NP$     = number of R&D projects
    $B$        = base evaluation cost as proportion of $V_p$ (%)

Equations 1, 2 and 3 can be solved simultaneously for Admin Cost to give:

$$AR = \frac{(B*V_i + y)}{V_i + y}$$

or $\quad y = \dfrac{V_i \cdot (AR - PSR_{in})}{1 - AR}$          ……… 3

At this point it is essential to calculate the number of successful projects, NSP, as follows:

$$NSP = NP * PSR \qquad \ldots.. \ 4$$

where $PSR = PSR_{in} + \Delta PSR$          ….. 5

and $\Delta PSR = f(y)$

The function f(y) can be expressed in several ways; in this analysis two alternative formulations are adopted, namely the linear function (Equation 6) or the logistics function (Equation 7) as follows:

$$f(y) = C.y \qquad \text{(for all C.y} < V_i \text{ and } \Delta PSR < \max \Delta PSR) \quad \ldots\ldots\ldots. \ 6$$

$$f(y) = C.(1 - e^{-k.y}) \qquad \text{(for C} = \max \Delta PSR) \quad \ldots\ldots\ldots. \ 7$$

By combining Equations 2, 4 and 5, we obtain:

$$NSP = (PSR_{in} + f(y)).(1 - AR).(V_p/V_i) \qquad \ldots\ldots\ldots \ 8$$

### 3.3 Step Three – the Portfolio Success Rate

In the final step, the portfolio success rate is calculated as follows:

$$PortSR = \frac{NSP}{(V_P/V_i)}$$

Or from Equation 8:

$$PortSR = (1 - AR).(PSR_{in} + f(y))$$

In most instances, the funder will be intent on maximising the portfolio success rate (PortSR), or the number of successful projects normalised for the theoretical maximum number of projects, and it is therefore important to consider how PortSR varies as a function of PSR and the other key input variables. Once this relationship is known, the optimum AR can be calculated. Mathematically this is normally done by expressing PortSR in terms of AR, differentiating the equation with respect to AR and then finding the roots of the equation in order to identify the peak of the curve (the point at which the gradient is zero). In the case of the linear function for y (Equation 3), a direct solution is possible as follows:

$$PortSR = (1 - AR).(PSR_{min} + C.y)$$
$$= AR.(C.V_i - PSR_{min}) + (PSR_{min} - C.V_i.B)$$

$$\frac{dPortSR}{dAR} = C.V_i - PSR_{min} \qquad \text{(for all C.y < Vi and ΔPSR < max ΔPSR)}$$

In this case, the $AR_{opt}$ is independent of y and depends only on the existing choice of $V_i$, C and $PSR_{min}$. The model predictions for the linear and logistics model as a function of AR are shown in Figure 3 and Figure 4 respectively.

**Figure 3. Linear model predictions for PortSR as a function of AR and $V_i$**

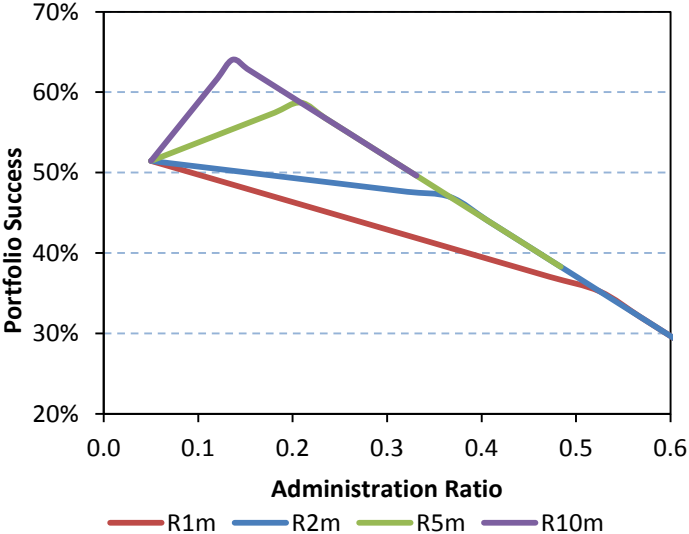

**Figure 4. Logistics model predictions for PortSR as a function of AR and $V_i$ (R million)**

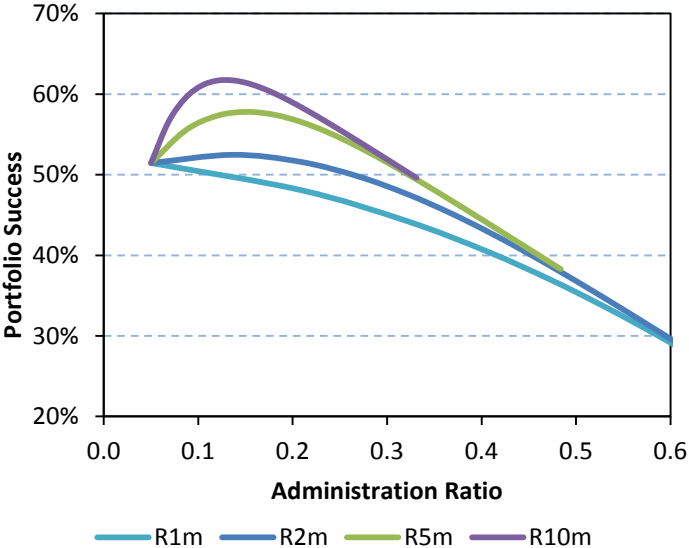

# 4. CALIBRATION AND TESTING OF THE MODEL

In this final section, the calibration of the model against the historical performance of THRIP is discussed. In particular, it is shown that the model correctly predicts THRIP performance over the period 2010 to 2014.

THRIP was established in 1992 as a funding instrument to support the development of 'high-level' technical skills for South African industry (9) and over the period 1994 to 2014, it has supported many hundreds of research projects at the universities and science councils. The THRIP model is based on the principles of a public-private partnership, with funding for each project being provided by both the industry partner and the Department of Trade and Industry, the latter to the amount of about $15 million per year (see Figure 5).

**Figure 5. Number of THRIP projects and total funds disbursed over the period 2000 to 2014 (funds adjusted to 2012 USD)**

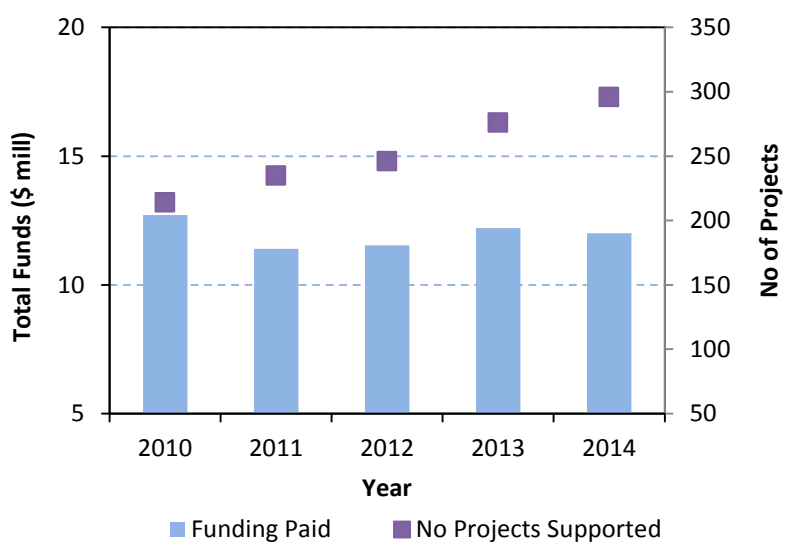

THRIP is administered by the National Research Foundation and assisted by an Advisory Council consisting of representatives from industry, government, higher education institutions, the Engineering Council of South Africa and the science councils. Information on many aspects of the fund such as the number of projects, the funds disbursed, administration costs and portfolio outcomes are published annually in the form of Annual Reports which can be downloaded from the programme's web page.

The programme is measured across a range of metrics including the numbers of researchers (undergraduate, graduate, post-graduate and permanent academic staff), supported projects, publications, patents, products, prototypes and small businesses. A study of the published data over the period 2010 to 2014 reveals two striking aspects of its performance. Although the total funds remained static as shown in Figure 5, the total number of funded per year increased with the result that the average funding per project ($V_i$) decreased by 23% and the overall administration costs increased by 22% (see Figure 6). Despite this decrease in funding per project, the outputs per project remained almost constant with the result that the overall portfolio performance increased by 35% (see Figure 7).

**Figure 6. THRIP funding per project (ZAR 2012) and the administration ratio**

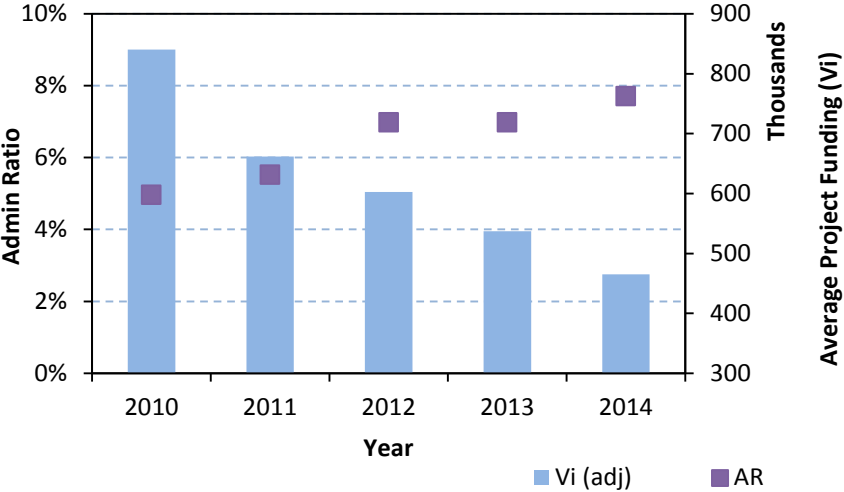

**Figure 7. THRIP performance and funding, normalised for 2010**

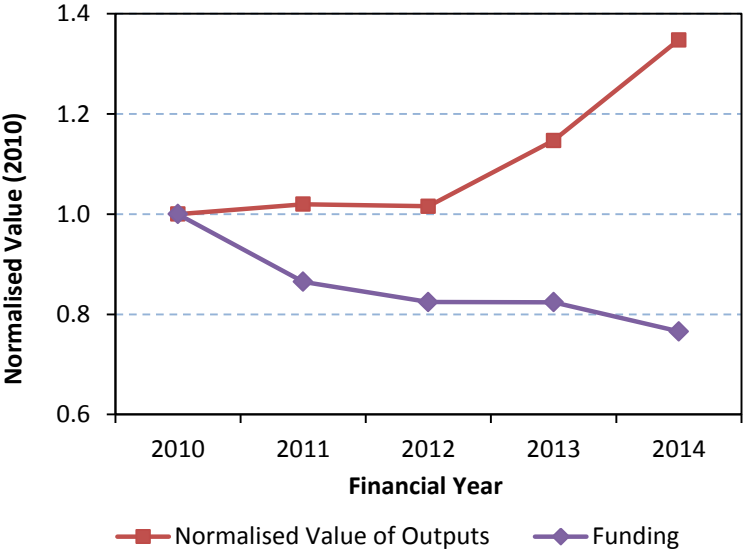

Normalisation of the outputs into a single metric (as shown in Figure 7) was undertaken using the following approach. Indicators of publications, patents, doctoral students and Masters students were selected as being representative of the most significant outputs from the programme. These outputs were then combined into a single metric based on the weighting factors of publications: Masters: Doctorate: Patents of 1:2:5:30. The latter values have been obtained from estimates for the relative monetary value of each output, with the base quantity being the value of a single publication set at $12,000 by the National Department of Education (16).

Analysis of the administration costs as a function of the number of projects suggests that the incremental cost per project is $3,240. This additional cost is more than offset by the additional output, with the result that the overall portfolio performance increases as the number of projects increase. Defining THRIP return on investment as the net value of the significant outputs divided by the total programme expenditure (disbursement and

administration), it is then apparent that the return has increased as the average project funding has decreased (see Figure 8).

**Figure 8. THRIP return on investment as a function of $V_i$**

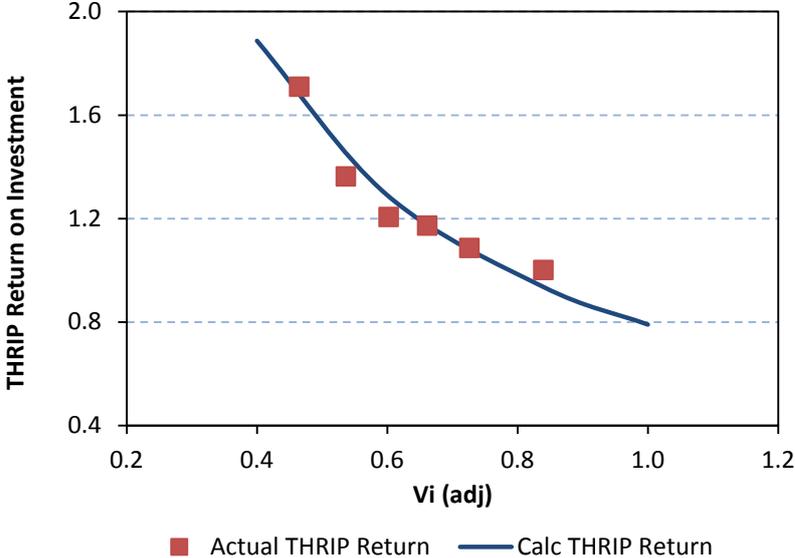

This result is surprising and counter-intuitive. Decreasing project funding should induce a decrease in project outputs, and increasing the administration burden should reduce the overall portfolio success rate. However the project outputs are insensitive to the level of funding within the values of the period 2010 to 2014, with the result that the portfolio performance increased even though more funds were being allocated to administration.

This result is closely predicted by the outputs of the funding model as described in the previous section and as shown in Figure 8. Confirmation of the actual with the predicted results provides a level of confidence in the model structure and initial data. However further work will be required to validate the model and in particular its calculation of the optimum administration ratio.

## 5. DISCUSSION

The objective of the research has been to provide a better understanding of administration costs in funding agencies, and in particular the relationship between portfolio success or performance and the administration ratio. A commonly expressed sentiment is that administration expenses add little value to portfolio outcomes and they should be kept as low as possible (13).

There is, however a counter-proposition which is core to this research, namely that a greater level of portfolio administration increases the success rate of projects up to a certain level of saturation. This proposition is considered as reasonable since in the absence of any proposal evaluation and screening or administration, there are no quality controls and a lower success rate can be expected. Furthermore evidence for the improvement in project outcomes as a consequence of more intensive supervision is reported in several areas, including by the business incubation community. For instance, the National Business Incubation Association of the USA reports that "after five years, businesses that were nurtured in a business incubator have a survival rate of 87 percent; in comparison, the survival rate for companies that go it alone without the benefits of incubator support is 44 percent" (17).

However it is apparent that the higher the administration cost, the lower the value of total project funding since both expenses are funded by $V_P$. As a consequence, it is expected that there will be a trade-off between AR and PSR. At some point, there will be an optimum AR, beyond which increased administration will only decrease the overall portfolio success rate where the latter is normalised to the total number of theoretical projects. The optimum value, however, is highly dependent on the characteristics of the fund, the agency efficiency and the expectations of the funder. The remainder of this analysis deals with the mathematical formulation of these relationships, and how the optimum AR can be determined.

In reality, efficient administration can add considerable value to a fund's performance by improving project screening (ex-ante evaluation), monitoring, reporting and ex-post evaluation. Although there is limited data at present on the relationship between the administration ratio and portfolio success rates, it is evident from other sectors that increased administration can in fact improve funding instrument outcomes. However this improvement has an associated cost and will reach a certain limit, beyond which increasing administration activity only serve to decrease the available funding for research projects without further improving the portfolio success.

The important question, and one which has been explored in this study, is to establish at what this optimum is achieved. In this research, a mathematical model has been constructed to enable the calculation of the optimum administration ratio and hence the optimum portfolio success rate. The model has been calibrated using historical data for THRIP and has then been used to explore the performance of THRIP over the period 2010 to 2014.

A curious feature of this performance is that the programme has increased its outputs (publications, patents and student qualifications) despite an increasing administration ratio and an almost static level of funding in real terms. The analysis reveals that the reason for the improvement is an increased number of funded projects per year. Even though the administration ratio has increased and the average funding per project has decreased, the individual projects have maintained the same performance output, leading to an overall increase of outputs by 35%.

This is a surprising and counter-intuitive result, and could not have been predicted without the insight of the model. It is a clear example of fund management where the portfolio success rate has increased with increasing administration costs and ratio. As noted in the introduction, the common perception is that the administration ratio should be minimised. In reality, a higher administration ratio may indeed be justifiable depending on whether the present cost lies below its optimum value.

The results of the THRIP analysis suggest that the model will have broader utility than just the THRIP analysis. Other insights from the model include the observations that in general, a portfolio of smaller projects (<$100,000) does not support a high evaluation cost since the administration ratio rapidly increases leading to a decrease in the number of funded projects; and conversely larger projects warrant a more detailed evaluation, leading to higher portfolio success rate. However an administration ratio of above 20% is rarely justifiable, as shown in Figure 4. Future work will focus on other funds and the collection of data from a wider range of instruments.

# 6. CONCLUSION

It is clear from the study that the value of administration within funding agencies can be misconceived. In particular, more intense, and hence higher cost, administration may be justifiable and will lead to an increase in both project success rates and overall portfolio success. Such an outcome has been observed for THRIP, which has improved its performance over the period 2010 to 2014 despite an increase in the administration ratio.

However this study is an initial investigation based on a model which has been calibrated with a limited dataset only. Further research is required in order to expand the data and validate the structure of the model against other agencies. It is apparent that a great deal can be done to improve the relationships between wholesale funders, portfolio managers, performance agencies and the research community. A more detailed and informed understanding of the administration ratio should form part of such an initiative.